\documentclass[preprint]{aastex631}

\usepackage{slantsc}
\usepackage{threeparttable}

\begin{document}

%% remember to change the title and authors
\title{The Kinematic and Chemical Properties of the Close-in Planet Host Star 8 UMi}

%% author information
\author[0009-0008-2988-2680]{Huiling Chen}
\affiliation{Department of Astronomy, School of Physics, Peking University, Beijing 100871, People's Republic of China}
\affiliation{Kavli Institute for Astronomy and Astrophysics, Peking University, Beijing 100871, People's Republic of China}

\author[0000-0003-3250-2876]{Yang Huang}
\affiliation{School of Astronomy and Space Science, University of Chinese Academy of Science, Beijing 100049, People's Republic of China}
\affiliation{CAS Key Laboratory of Optical Astronomy, National Astronomical Observatories, Chinese Academy of Sciences, Beijing, 100101, People's Republic of China}

\author[0000-0003-4027-4711]{Wei Zhu}
\affiliation{Department of Astronomy, Tsinghua University, Beijing 100084, People's Republic of China}

\author[0000-0003-4573-6233]{Timothy C. Beers}
\affiliation{Department of Physics and Astronomy, University of Notre Dame, Notre Dame, IN 46556, USA}
\affiliation{Joint Institute for Nuclear Astrophysics -- Center for the Evolution of the Elements (JINA-CEE), USA}

\author{Renjing Xie}
\affiliation{CAS Key Laboratory of Optical Astronomy, National Astronomical Observatories, Chinese Academy of Sciences, Beijing, 100101, People's Republic of China}

\author[0000-0002-4391-2822]{Yutao Zhou}
\affiliation{College of Mathematics and Physics, Guangxi Minzu University, Nanning 530006, People's Republic of China}

\author[0000-0002-6937-9034]{Sharon Xuesong Wang}
\affiliation{Department of Astronomy, Tsinghua University, Beijing 100084, People's Republic of China}

\author[0000-0002-9702-4441]{Wei Wang}
\affiliation{CAS Key Laboratory of Optical Astronomy, National Astronomical Observatories, Chinese Academy of Sciences, Beijing, 100101, People's Republic of China}

\author[0000-0002-8709-4665]{Sofya Alexeeva}
\affiliation{CAS Key Laboratory of Optical Astronomy, National Astronomical Observatories, Chinese Academy of Sciences, Beijing, 100101, People's Republic of China}

\author{Qikang Feng}
\affiliation{Department of Astronomy, School of Physics, Peking University, Beijing 100871, People's Republic of China}
\affiliation{Kavli Institute for Astronomy and Astrophysics, Peking University, Beijing 100871, People's Republic of China}

\author{Haozhu Fu}
\affiliation{Department of Astronomy, School of Physics, Peking University, Beijing 100871, People's Republic of China}
\affiliation{Kavli Institute for Astronomy and Astrophysics, Peking University, Beijing 100871, People's Republic of China}

\author[0000-0002-0389-9264]{Haining Li}
\affiliation{CAS Key Laboratory of Optical Astronomy, National Astronomical Observatories, Chinese Academy of Sciences, Beijing, 100101, People's Republic of China}

\author[0000-0002-6540-7042]{Lile Wang}
\affiliation{Department of Astronomy, School of Physics, Peking University, Beijing 100871, People's Republic of China}
\affiliation{Kavli Institute for Astronomy and Astrophysics, Peking University, Beijing 100871, People's Republic of China}

\author[0000-0002-7727-1699]{Huawei Zhang}
\affiliation{Department of Astronomy, School of Physics, Peking University, Beijing 100871, People's Republic of China}
\affiliation{Kavli Institute for Astronomy and Astrophysics, Peking University, Beijing 100871, People's Republic of China}

\correspondingauthor{Yang Huang and Huawei Zhang}
\email{huangyang@ucas.ac.cn and zhanghw@pku.edu.cn}

%% abstract should be no more than 250 words
\begin{abstract}
A recent study by Hon et al.\ reported that a close-in planet around the red clump star, 8\,UMi, should have been engulfed during the expansion phase of its parent star's evolution.
They explained the survival of this exoplanet through a binary-merger channel for 8\,UMi.
The key to testing this formation scenario is to derive the true age of this star: is it an old "imposter" resulting from a binary merger, or a genuinely young red clump giant?
To accomplish this, we derive kinematic and chemical properties for 8\,UMi using astrometric data from {\it Gaia} DR3 and the element-abundance pattern measured from a high-resolution ($R \sim 75,000$) spectrum taken by SOPHIE. 
Our analysis shows that 8\,UMi is a normal thin-disk star with orbital rotation speed of $\it{V}_\mathrm{\phi}=\mathrm{244.96 \, km s^{-1}}$, and possesses a Solar metallicity ([Fe/H]\,$= -0.05 \pm 0.07$) and $\alpha$-element abundance ratio ([$\alpha$/Fe]\,$= +0.01 \pm 0.03$). 
By adopting well-established relationships between age and space 
velocities/elemental abundances, we estimate a kinematic age of $3.50^{+3.00}_{-2.00}$\,Gyr, and a chemical age of $3.25^{+2.50}_{-1.50}$\,Gyr from [C/N] and $3.47 \pm 1.96$\,Gyr from [Y/Mg] for 8\,UMi, respectively.
These estimates are consistent with the isochrone-fitting age ($1.90^{+1.15}_{-0.30}$\,Gyr) of 8\,UMi, but are all much younger than the timescale required in a binary-merger scenario.
This result challenges the binary-merger model; the existence of such a closely orbiting exoplanet around a giant star remains a mystery yet to be resolved.
\end{abstract}

%% abstract has been checked by ChatGPT
%% Keywords should appear after the \end{abstract} command.
%% The AAS Journals now uses Unified Astronomy Thesaurus concepts:
%% https://astrothesaurus.org
%% You will be asked to selected these concepts during the submission process
%% but this old "keyword" functionality is maintained in case authors want
%% to include these concepts in their preprints.

\section{Introduction}
%\label{sec:intro}
Since the first detections of exoplanets \citep{1992Natur.355..145W,1995Natur.378..355M}, over 5,000 exoplanets have been discovered\footnote{\url{https://exoplanetarchive.ipac.caltech.edu/}}. 
%\citep{Akeson:2013}
This extensive sample has revolutionized our knowledge of the formation and evolution of planets \citep[e.g.,][]{2021ARA&A..59..291Z}. 
While the majority of confirmed exoplanets are found around main-sequence stars, planets around evolved stars (see \citealt{Dollinger:2021} and references therein), or even around stellar remnants \citep[e.g.,][]{1992Natur.355..145W}, have also been detected. These exotic planets provide useful constraints on the long-term evolution of planetary systems.
% To me the following text is not much relevant to the present discussion, so I have removed it. But I'm fine if you want to include it back.

The planetary system around the red giant 8\,UMi was first discovered by \citet{2015A&A...584A..79L}. 
The mass and eccentricity of the planet, Halla, were determined using the radial-velocity method with high-resolution spectra acquired with the fiber-fed Bohyunsan Observatory Echelle Spectrograph (BOES) mounted on the 1.8-m telescope at Bohyunsan Optical Astronomy Observatory (BOAO).
The atmospheric parameters of the host star were also carefully derived from these high-resolution spectra. This discovery is further confirmed by \citet{2023Natur.618..917H}, based on 135 additional radial-velocity measurements using the HIRES spectrograph installed on the Keck-I telescope on Mauna Kea, Hawaii.
The semi-major axis of Halla is only 0.462$\pm$0.006\,au, calculated from fits to its radial-velocity curve; its mass has been estimated from asteroseismology.
The measured period spacing $\Delta \Pi$ from asteroseismology suggests that 8\,UMi is a helium-burning red clump giant star \citep{2018ApJS..236...42Y}. 
Given these properties, Halla should have been engulfed and destroyed by 8\,UMi during the penultimate phase of its evolution, the tip of the red giant branch, when the star expanded out to 0.7\,au.
To avoid the engulfment of Halla, \citet{2023Natur.618..917H} proposed a binary-merger formation scenario for 8\,UMi. 
According to their binary evolution model, a close binary composed of two lower-mass stars was required to form a 8\,UMi-like red clump star. 
If this scenario were correct, 8\,UMi would be old ($\sim8.6\,$Gyr according to the binary-merger model in \citealt{2023Natur.618..917H}), and inherit the kinematic and chemical properties from its progenitor binary system. It should belong to the Milky Way's thick-disk population, given the evolution timescale required in the binary-merger model.

In order to check the aforementioned binary-merger scenario, it is crucial to constrain the age of the system.
In this letter, the kinematic and chemical properties of the host star 8\,UMi are carefully investigated using astrometric data from {\it Gaia} DR3 \citep{2023A&A...674A...1G} and a high-resolution ($R \sim 75,000$) spectrum from the SOPHIE archive\footnote{\url{http://atlas/obs-hp.fr/sophie/}}. 
This information is used to derive the kinematic and chemical ages of 8\,UMi, based on well-established relations between age and space velocities/chemical elemental-abundance ratios.
In Section\,\ref{sec:data}, we briefly introduce the adopted data and the analysis of high-resolution spectrum. 
Section\,\ref{Results and Discussion} presents the mass, age, measurements of 3D positions, 3D space velocities, and chemical-abundance pattern of 8\,UMi. Kinematic and chemical constraints on the true age of 8\,UMi are also provided. Based on these results, we conclude that the binary-merger scenario is unlikely, and that alternative explanations must be considered for the 8\,UMi system.  
Section\,\ref{sec:conclusion} presents a brief summary.

\vskip 2cm
\section{Data and Spectrum Analysis} \label{sec:data}

\subsection{Gaia DR3} \label{sec:Gaia Observation}

In this study, the astrometric parameters and the broad-band photometry of 8\,UMi are retrieved from {\it Gaia} DR3\footnote{\url{https://gea.esac.esa.int/archive/}} \citep{2023A&A...674A...1G}. The derived distance of 8\,UMi is $163.7 \pm 0.4$\,pc, measured from the {\it Gaia} parallax with the zero-point corrected.
To correct for the reddening, the value of extinction $E (B-V) = 0.02$ is taken from the reddening map of \citet{1998ApJ...500..525S}, with systematics corrected. 
The extinction coefficients of the three {\it Gaia} bands are taken from Table\,1 of \citet{2021ApJ...907...68H}.
The intrinsic color and absolute magnitude of 8\,UMi are thus $(B_{\rm P} - R_{\rm P})_0 = 1.13 \pm 0.03$ and $M_{G} = 0.45 \pm 0.05$, respectively.
The uncertainties are from the photometric errors and the reddening correction by assuming a constant error of 0.02 in $E (B-V)$.

\subsection{High-resolution Spectrum and Analysis} \label{sec:HRS Observation}

Analysis of a high-resolution spectrum is required to accurately infer the elemental-abundance ratios of 8\,UMi.
We searched for observation records in public spectral data archives, and retrieved the high-resolution spectrum of 8\,UMi from the SOPHIE archive. This spectrum was obtained with the 1.93-m telescope at Haute Provence Observatory equipped with the SOPHIE {\'e}chelle spectrograph.
The downloaded spectrum is processed through the standard data reduction pipeline of SOPHIE. First, a 
two-dimensional spectrum (E2DS) is obtained after optimal order extraction, cosmic-ray extraction, wavelength calibration, and spectral flat-field correction to the raw data. 
Secondly, the pipeline merges the 39 spectral orders after correction of the blaze function, yielding a one-dimensional spectrum \citep[S1D;][]{2009A&A...505..853B}. 
The observation mode employed for 8\,UMi was the high-resolution mode (HR mode), with a spectral resolution of 
$R\sim$75,000, covering the wavelength range from 3872 to 6943\,{\AA}. 
The signal-to-noise ratio (SNR) of this spectrum is generally above 100 (peak SNR $\sim$ 350 at around 6300\,{\AA}), except at the very blue edge ($\le 4050$\,{\AA}), which is excluded in the following analysis.

For the downloaded spectrum, we use the cross-correlation function method to calculate its line-of-sight velocity $v_{\rm los} =-9.58\pm 0.03$\,km\,s$^{-1}$. 
The spectrum is corrected to the rest-frame using the derived $v_{\rm los}$, and then is normalized locally by division with a pseudo-continuum defined manually. To derive atmospheric parameters (i.e., effective temperature $T_{\rm eff}$, surface gravity log\,$g$, and metallicity [Fe/H]) of 8\,UMi, we employ the equivalent-width method with the MOOG radiative transfer code \citep{2012ascl.soft02009S}. The MARCS atmospheric model \citep{2008A&A...486..951G}, the linelist compiled from \citet{2000A&AS..141..491C} and \citet{2015ApJ...808..148S}, and the Solar abundance of \citet{2007SSRv..130..105G} are adopted.
All of these processing steps are combined in the integrated software package {\sc i}S{\sc pec} \citep{2019MNRAS.486.2075B}.
The equivalent widths (EWs) of the atomic iron lines are measured, based on the compiled linelist mentioned above.
Specifically, a Gaussian function is adopted to fit unblended single atomic absorption lines.
Ultimately, the EWs of 30 Fe\,{\sc i} lines and 7 Fe\,{\sc ii} well-separated and weak to mild-strength lines (20 to 100\,m\AA) are measured in this way. 

\begin{figure}[ht!]
\centering
\includegraphics[scale=0.45]{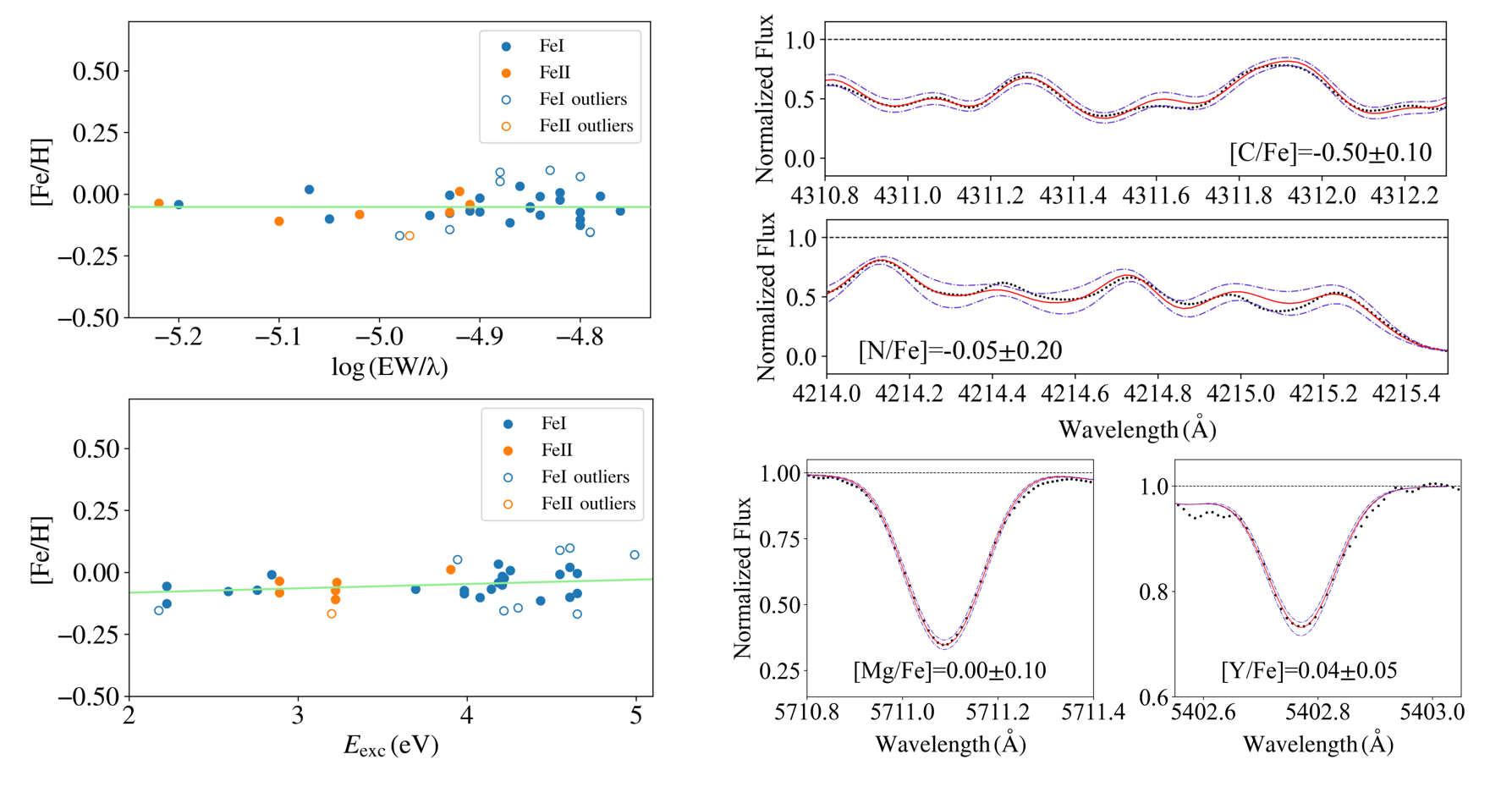}
\caption{Left panels: [Fe/H] estimates as a function of reduced equivalent width (upper sub-panel) and excitation potential (lower sub-panel). The open circles represent the lines filtered by the sigma-clipping process of {\sc i}S{\sc pec} with a 0.90 weight scale. Right panels: Synthetic spectrum-fitting examples (red solid line) of 8\,UMi in the abundance-determination process. The black dots show the observed spectrum. The upper two sub-panels show portions of the CH and CN bands with 1-$\sigma$ errors (blue dash-dot line). The lower two sub-panels show the fitting examples for the Mg 5711\,$\mathrm{\AA}$ and Y 5402\,$\mathrm{\AA}$ lines, with abundance variations for 0.10\,dex and 0.05\,dex, respectively.}
\label{fig:chemical pattern}
\end{figure}

The measured EWs, using $T_{\rm eff} = 4847$\,K, [Fe/H]\,$= -0.03$, and $v_{\rm mic} = 1.30$\,km\,s$^{-1}$ from \citet{2023Natur.618..917H} as an initial guess, are employed to derive the atmospheric parameters of 8\,UMi using the nonlinear least-squares fitting algorithm installed in {\sc i}S{\sc pec}, which considers both excitation equilibrium and ionization balances.
For surface gravity, we have two estimates: one, log\,$g$ = 2.57, is from the asteroseismologic scaling relation \citep{2018A&A...616A.104K} with maximum power $\nu_{\rm max}$ measured by \citet{2023Natur.618..917H} from TESS, with a correction for a $\sim$4\% systematic offset found in \citet{2022MNRAS.512.1677S}; another, $\log g$ = 2.63, is from isochrone-fitting based on the Bayesian technique (see Sec.\,\ref{Mass and Age From Isochrone Fitting} for details).
A mean value of $\log g$ = 2.60 is the input value in {\sc i}S{\sc pec} for log\,$g$; it changes slightly with the changes of $T_{\rm eff}$ and [Fe/H] during the iteration.
Finally, the atmospheric parameters are found to be: 
$T_{\rm eff} = 4897 \pm 100 $\,K, log\,$g = 2.60 \pm 0.07$, [Fe/H]\,$= -0.05 \pm 0.07$, and $v_{\rm mic} = 1.48 \pm 0.10$\,km\,s$^{-1}$. Fig.\,\ref{fig:chemical pattern} shows the [Fe/H] estimates as a function of reduced EWs and excitation potential. The slopes between [Fe\,I\&II/H] and reduced equivalent width/excitation potential are both flat\footnote{For excitation balance, a not significant slope of 0.015 dex eV$^{-1}$ is detected.}.
%There is no slope between the estimates of [Fe/H] and reduced EWs, while a small slope (0.015 dex eV$^{-1}$) exists between estimated [Fe/H] and excitation potential because of the $\log g$ prior in the iteration. Errors from this small slope are added to the error estimate of $T_{\rm eff}$. 
The difference between [FeI/H] and [FeII/H] is smaller than 0.01\,dex. In general, the derived atmospheric parameters are consistent with those from \citet{2023Natur.618..917H}, as well as those estimated in the original discovery paper \citep{2015A&A...584A..79L}.

Based on the estimated atmospheric parameters, the chemical abundances for a total of 23 elements are measured using the software packages SIU \citep{1991PhDThesis} and TAME \citep{2012MNRAS.425.3162K} by the spectrum-synthesis method with the assumption of LTE. A differential analysis is adopted when deriving abundances. The original atomic parameters are from different sources for different elements. The atomic parameters for the CH molecular lines are taken from \citet{2014A&A...571A..47M}. 
The linelist is generated using LINEMAKE \citep{2021RNAAS...5...92P} around 4215\,$\mathrm{\AA}$ for deriving the abundance of N. 
For other elements, the atomic parameters are drawn from several studies including \citet{2009A&A...497..563N}, \citet{2016ApJ...833..225Z}, and \citet{2022MNRAS.515.1510S}. 
We refine the $\log gf$ values of each atomic line to make sure they well-reproduce atomic absorption-line profiles in the Solar spectrum of \citet{1984sfat.book.....K}. The refined $\log gf$ values, excitation potentials, EWs observed in the Solar spectrum, EWs measured in 8 UMi, and the corresponding abundances calculated for 8 UMi of individual lines are presented in Table\,\ref{Tab:line data}. Synthetic line-fitting examples for CH, CN, Mg and Y in 8\,UMi are shown in Fig.\,\ref{fig:chemical pattern}. For the Solar spectrum, the synthetic line-fitting results are presented in Fig.\,\ref{fig:solar_fitting}.
%For Solar spectrum, the synthetic line fitting results of Mg and Y are shown in appendix \ref{Fitting examples in Solar spectrum}. Since the molecular bands for CH and CN have different sensitivity in giant and dwarf spectrum, we do not present the fitting of CH and CN in Solar spectrum here. 

\vskip 1.5cm
\section{Results and Discussion} \label{Results and Discussion}

\begin{figure}[ht!]
\centering
\includegraphics[scale=0.55]{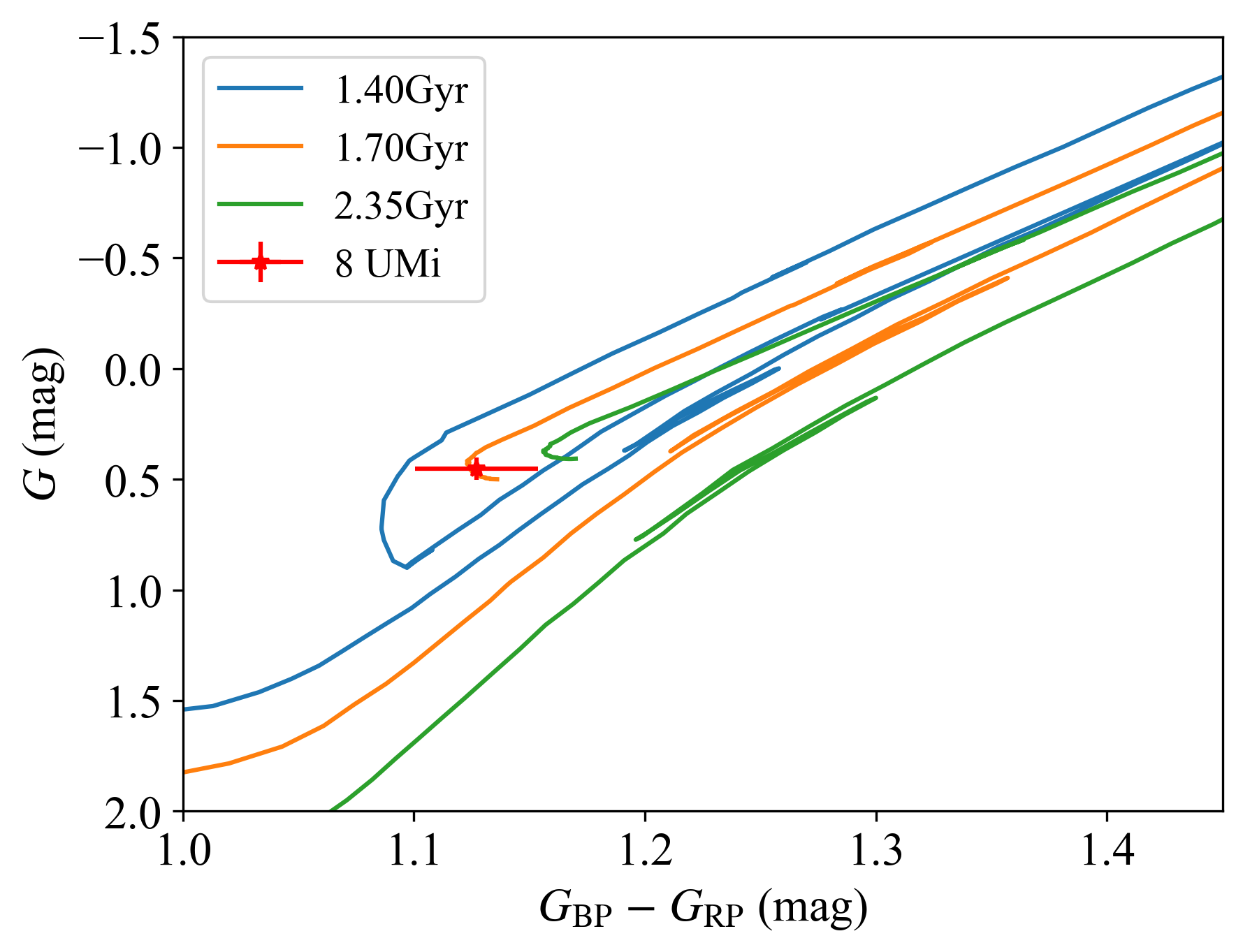}
\includegraphics[scale=0.55]{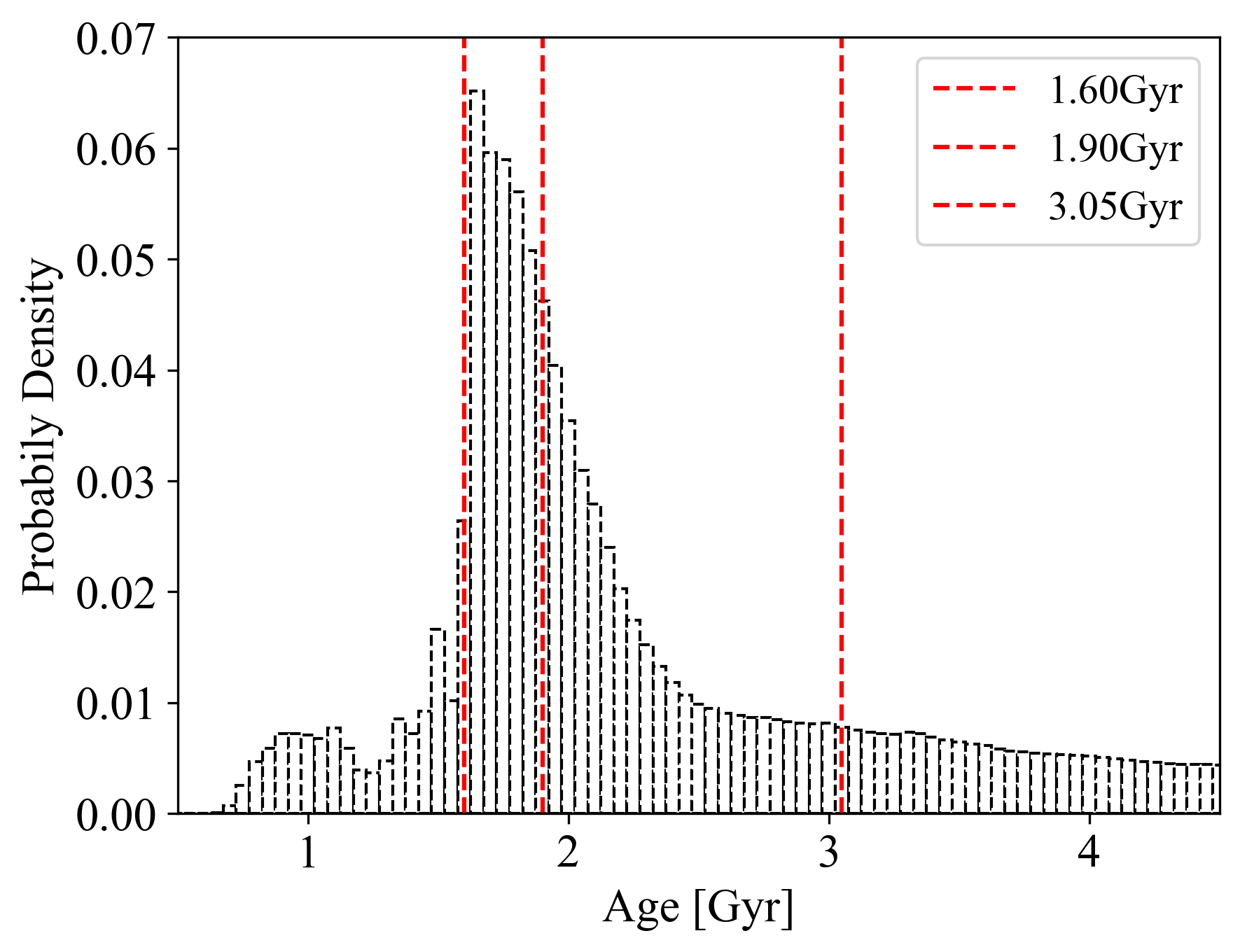}
\caption{Left panel: Theoretical isochrones of 1.40\,Gyr, 1.70\,Gyr, and 2.35\,Gyr (from left to right) and [M/H]=$-0.05$ in the $(B_{\rm P} - R_{\rm P})_0$ -- $M_{\rm G}$ diagram. 
The red star with error bars represents 8\,UMi. Right panel: The posterior probability density distribution of age calculated by our Bayesian method (see Sec.\,\ref{Mass and Age From Isochrone Fitting}). 
The red-dashed lines of 1.60\,Gyr, 1.90\,Gyr, and 3.05\,Gyr  represent the 16\%, 50\% and 84\% cumulative probability levels.}
\label{fig:isochrone_age}
\end{figure}

\subsection{Mass and Age from Isochrone Fitting}  \label{Mass and Age From Isochrone Fitting}

To derive the mass and age of 8\,UMi, we adopt a conventional Bayesian method, similar to that developed in \citet{2022ApJ...925..164H}, by matching the observed stellar parameters with theoretical isochrones. 
The theoretical isochrones are taken from PARSEC \citep{2012MNRAS.427..127B}, with [M/H] from $-$0.12 to +0.02 in steps of 0.01\,dex, and age from 0.5 to 4.5\,Gyr in steps of 0.05\,Gyr.
The full grids include over 540,000 individual models.
The input observable constraints are intrinsic color $(B_{\rm P} - R_{\rm P})_0$ and absolute magnitude $M_{G}$ (see Sec.\,\ref{sec:Gaia Observation}).
As an example, Fig.\,2 compares 8\,UMi to model isochrones 
in the $M_G$--$(G_{\rm BP} - G_{\rm RP})_0$ diagram (left panel), and presents the posterior probability distribution function (PDF) of age yielded by the Bayesian estimate (right panel).
Based on the PDF, the age of 8\,UMi is found to be $\tau = 1.9^{+1.2}_{-0.3}$\,Gyr. 
The upper and lower uncertainties correspond to the 16\% and 84\% percentiles of the resulting PDF.
This suggests 8\,UMi is a young star, if it has evolved as a single star.

The isochrone fitting yields a mass of $1.70^{+0.15}_{-0.25}\, M_{\odot}$ for 8 UMi. This is consistent with the stellar mass derived in a similar way by the original planet discovery paper \citep{2015A&A...584A..79L}, but larger by 13\% than the asteroseismic measurement ($1.51 \pm 0.05 \,M_\odot$) of \citet{2023Natur.618..917H}. We do not attempt to resolve this discrepancy here, but note that the asteroseismic measurements on 8 UMi by two different studies, \citet{2023Natur.618..917H} and \citet{Hatt:2023}, are inconsistent with each other in the oscillation frequency at the maximum power, $\nu_{\rm max}$, which is the key parameter that determines the stellar mass. These two studies both used TESS data, although they differ in the number of sectors (8 vs.\ 12). According to \citet{Hatt:2023}, the oscillation frequency at the maximum power, $\nu_{\rm max}$, is determined to be $50.8 \pm 2.4\,\mu$Hz, which would suggest a host mass of about $2\,M_\odot$.

The value of the host star mass may be the key to resolve the puzzle of 8\,UMi b. A higher host mass means that the planetary orbit is slightly wider at a given orbital period. In the mass range 1.5--2.0$\,M_\odot$, a higher mass also produces a smaller stellar size at the tip of red giant branch. The latter is best illustrated in the panel (b) of Figure~3 in \citet{2023Natur.618..917H} for the case of the 8\,UMi system. As a consequence, the impact of the tidal effect also sensitively depends on the host star mass, and a slightly higher mass can reduce the chance of planet engulfment during the RGB and Helium-flash phases, as theoretical studies have shown \citep[e.g.,][]{2011ApJ...737...66K}.
%tidal effects are considered during the the host star's RGB and helium-flash evolution, planets originally located on farther orbits can be pulled closer to their host star \citep{2011ApJ...737...66K}. This phenomenon can result in planets orbiting at 0.5\,au. for lower-mass stars (e.g,\,$1.8\,M_\odot$ in Fig.\,1 of \citealt{2011ApJ...737...66K}).

%A mass of $\sim1.7\,M_\odot$ for 8\,UMi means that the planet may have never been engulfed in the single-star evolution scenario, at least according to some stellar evolution models.
%One possible reason is that the $\mathrm{\nu_{max}}$ measured from TESS is underestimated by 3-5\%, at $\mathrm{\nu_{max}}$ between 45 and 50\,$\mathrm{\mu}$Hz, as compared to those accurately measured from Kepler \citep{2022MNRAS.512.1677S}.
%If correcting for this systematic offset, the asteroseismology mass would be 1.68\,$M_{\odot}$ (assuming a 3\% offset)  to 1.78\,$M_{\odot}$ (assuming a 5\% offset), fully consistent with the isochrone-fitting mass.
%Finally, as a by-product, the surface gravity is estimated to be log\,$g$ = $2.65 \pm 0.07$, again using the Bayesian technique.

%\vskip 1cm
\subsection{Space Velocities and Kinematic Age}   \label{3D Space Velocities and Kinematic Age}

\begin{figure}[ht!]
\centering
\includegraphics[scale=0.53]{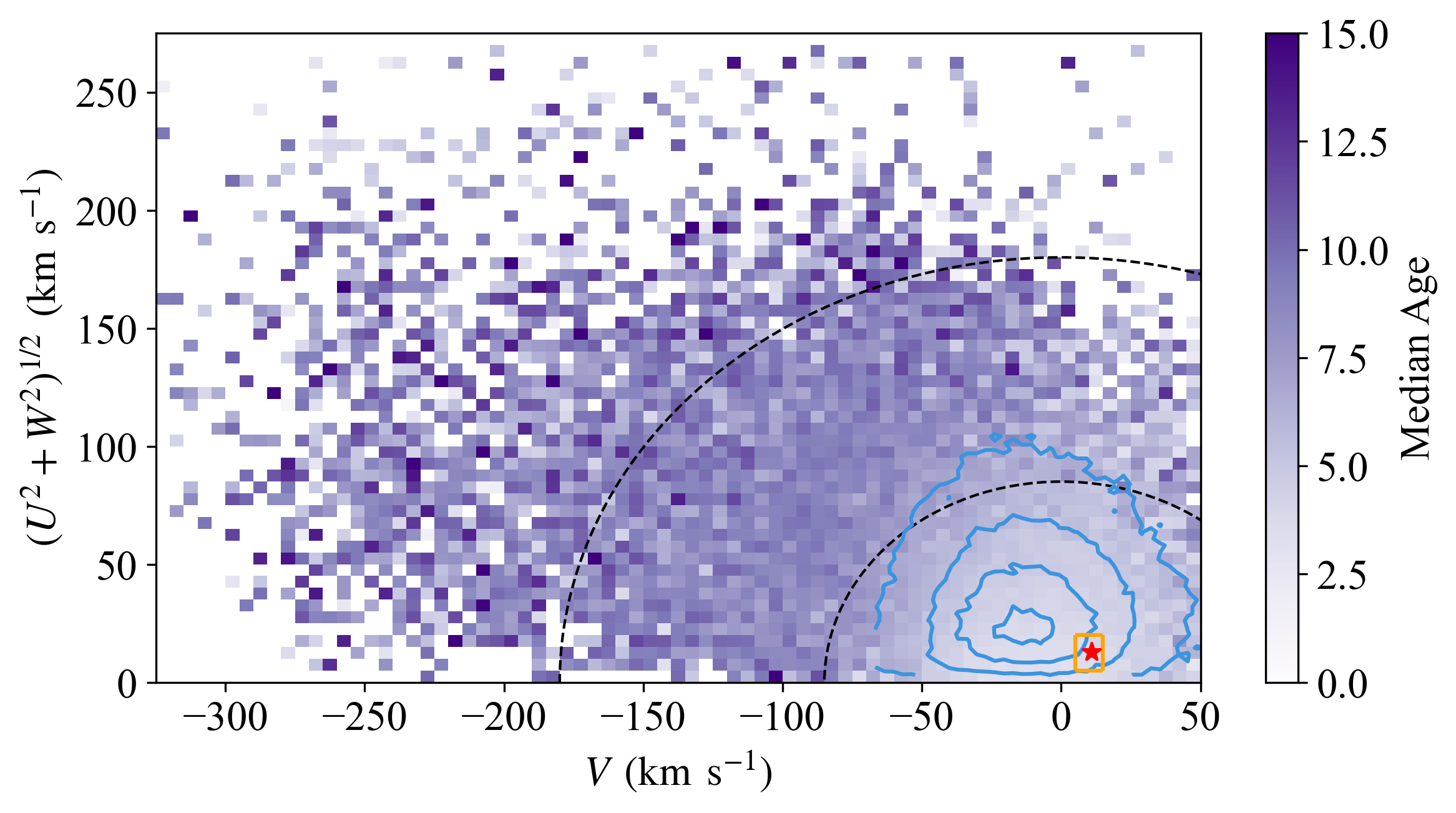}
\includegraphics[scale=0.50]{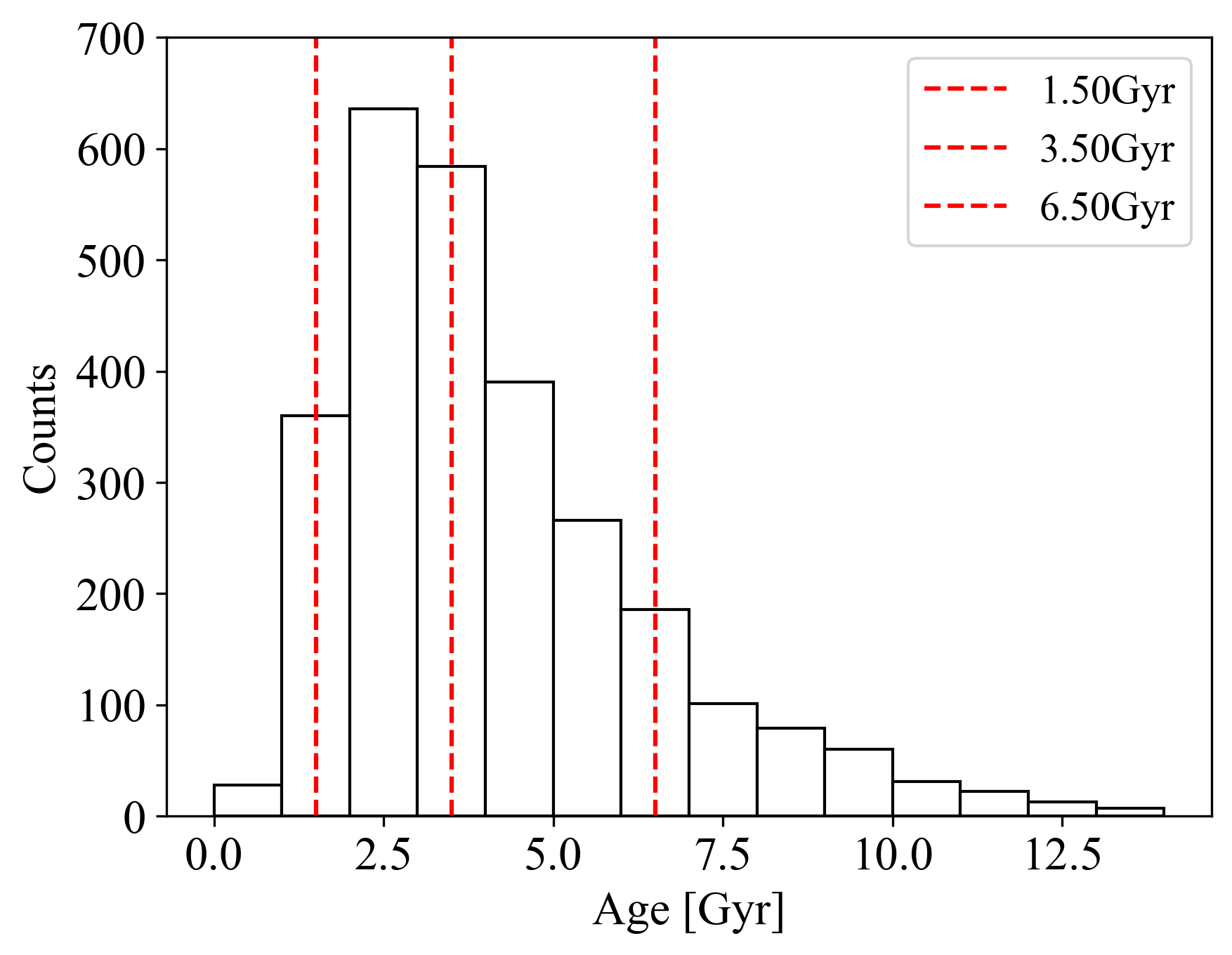}
\caption{Left panel: Toomre Diagram for 8\,UMi (red star). Overplotted background is the distribution of field red clump stars from \citet{2020ApJS..249...29H}, color coded by median age with the color bar shown on the right side. 
The bin size is 5 km\,s$^{-1}$ and 5 km\,s$^{-1}$ in the $X$-axis and $Y$-axis, respectively.
The two dashed circles delineate constant total space velocities with respect to the local standard of rest of $\it{V}_\mathrm{tot}$ = 85 and 180 \,km\,s$^{-1}$, respectively, as commonly used to define a sample of thick-disk stars. The blue contours represent the 10\%, 50\%, 84\% and 99\% of the thin-disk population selected from the field red clump stars, based on the criteria taken from \citet{2021ApJ...909..115C}. 
The red rectangle ($V$ from 5 to 15\,km\,s$^{-1}$ and $\sqrt{U^2 + W^2}$ from 5 to 20\,km\,s$^{-1}$) selects stars with similar 3D space velocities to 8\,UMi.
Right panel: The age distribution of the red clump stars selected by the red rectangle shown in the left panel. Lines of 1.50\,Gyr, 3.50\,Gyr and 6.50\,Gyr mark the 16\%, 50\% and 84\% 
cumulative probabilities, respectively.}
\label{fig:kinematic age}
\end{figure}

The {\it Gaia} DR3 astrometric information and distance estimate (see Sec.\,\ref{sec:Gaia Observation}), as well as the radial velocity measured from the SOPHIE spectrum (see Sec.\,\ref{sec:HRS Observation}), are adopted to derive space velocities with the {\tt galpy} python package \citep{2015ApJS..216...29B}. 
For this calculation, the Solar motions are set to be ($U_\odot$, $V_\odot$, $W_\odot$)\,$= (-7.01, 10.13, 4.95)$\,km\,s$^{-1}$ \citep{2015MNRAS.449..162H}, and the circular speed at the Solar position set to be $V_{c} (R_0) = 234.04$\,km\,s$^{-1}$ from \citet{2023ApJ...946...73Z}.
For the Solar positions, we adopt a value of $R_0 = 8.178$\,kpc \citep{2019A&A...625L..10G} for the Galactocentric
distance, and a value of $Z_0 = 25$\,pc \citep{2016ARA&A..54..529B} for the vertical offset.
The resulting space velocity components of 8\,UMi are: 
($U$, $V$, $W$)\,$= (-10.28 \pm 0.20 , 10.92 \pm 0.12 , -7.61 \pm 0.09)$\,km\,s$^{-1}$ in a Cartesian coordinate system centered on the Sun and ($V_{R}$, $V_{\phi}$, $V_{Z}$)\,$= (-10.28 \pm 0.20 , 244.96 \pm 0.19,  -7.61 \pm 0.09)$\,km\,s$^{-1}$ in  
Galactocentric cylindrical coordinates. 
The three-dimensional positions of 8\,UMi are $R = 8.228 \pm 0.035$ \,kpc, $\phi = 0.809 \pm 0.004^{\circ} $ and $Z = 129.7\pm 0.5 $\,pc.
Errors in the space velocities and positions are calculated by 20,000 Monte Carlo (MC) simulations by sampling the measurement errors (assuming Gaussian distributions).
Fig.\,\ref{fig:kinematic age}a shows the location of 8\,UMi in the Toomre Diagram.
For comparison, the background shows the distribution of field red clump giant stars \citep[sample taken from][]{2020ApJS..249...29H} color coded by median age.
In \citet{2020ApJS..249...29H}, the ages of these red clump stars are estimated from LAMOST spectra,  using machine-learning techniques, with the ages calibrated using asteroseismic masses from {\it Kepler}.
The typical age uncertainty is 20\%.
In this figure, we also show density contours of the thin-disk population selected from the field red clump sample with the criteria from \citet{2021ApJ...909..115C}.
8\,UMi falls on the typical position of members of the thin-disk population with young median ages.

To provide a quantitative estimate of the kinematic age for 8\,UMi, red clump stars with 3D space velocities similar to 8\,UMi (i.e., $V$ between 5 and 15 km\,s$^{-1}$ and $\sqrt{U^2 + W^2}$ between 5 and 20 km\,s$^{-1}$) are selected. 
The right panel of Fig.\,\ref{fig:kinematic age} shows the age distribution of those selected red clump stars. 
This leads to an kinematic age estimate $\tau_{\rm kinematic}=3.5^{+3.0}_{-2.0}$\,Gyr for 8\,UMi.

\vskip 1cm
\subsection{Elemental-abundance Pattern and Chemical Age}  \label{Element-abundance Patterns and Chemical Age}

Table\,\ref{Tab:abundance} presents the chemical-abundance pattern for 8\,UMi determined using the methods described in Sec.\,\ref{sec:HRS Observation}. The errors of the derived abundances are also provided in Table\,\ref{Tab:abundance}. The values of $\sigma$ represent the statistical uncertainties in measurements, while the $\delta$[X/Fe] represents the systematic error estimated from changes of abundances by individually varying $T_{\rm eff}$, log\,$g $, [Fe/H], and $v_{\rm mic}$ according to the 1-$\sigma$ parameter uncertainties. Overall, the chemical results indicate that 8\,UMi, with $\mathrm{[Fe/H]=-0.05\pm0.07}$ and $\mathrm{[\alpha/Fe] = +0.01\pm0.03}$ is a typical thin-disk star. 
In addition,  four intriguing features are also seen in the chemical-abundance pattern of 8\,UMi. 
First, the lithium abundance is significantly enhanced, with $\mathrm{A(Li)\approx1.94}$, which is consistent with previous measurements of $\mathrm{A(Li)=2.0\pm0.2}$ from \citet{2011ApJ...730L..12K} and \citet{2020A&A...633A..34C}. 
Possible origins of this enhanced Li abundance are discussed in the next section.
Secondly, the carbon abundance is relatively low for such a Solar metallicity star. 
It may be attributed to extra mixing through the CNO cycle (aka the first dredge-up process) during the red giant branch (RGB) phase \citep[e.g.,][]{1999ApJ...510..232B}. 
Thirdly, $\mathrm{[Zn/Fe]=-0.29}$ is slightly lower than the Solar value. 
Note that \citet{2022MNRAS.515.1510S} reported a median $\mathrm{[Zn/Fe]\approx-0.10}$, with a lower boundary extending as low as $-0.30$, for thin-disk stars with 
Solar metallicity.
This suggests that sub-Solar zinc abundances are quite common for Solar-metallicity stars.
Finally, the $s$-process elements (Ba, La, Nd) exhibit sightly higher abundances, with an average value of $\mathrm{[X/Fe]\approx +0.20 }$ to $+0.30$. 
This super-Solar abundance is additional evidence for the genuinely young age for 8\,UMi, since the fraction of $s$-process elements increases with time in the Universe \citep{2018MNRAS.474.2580S}.
%This suggests that 8\,UMi may have been born in a cloud that was enriched by an asymptotic giant branch (AGB) star \citep{2001ApJ...557..802B}.

\begin{figure}[ht!]
\centering
\includegraphics[scale=0.58]{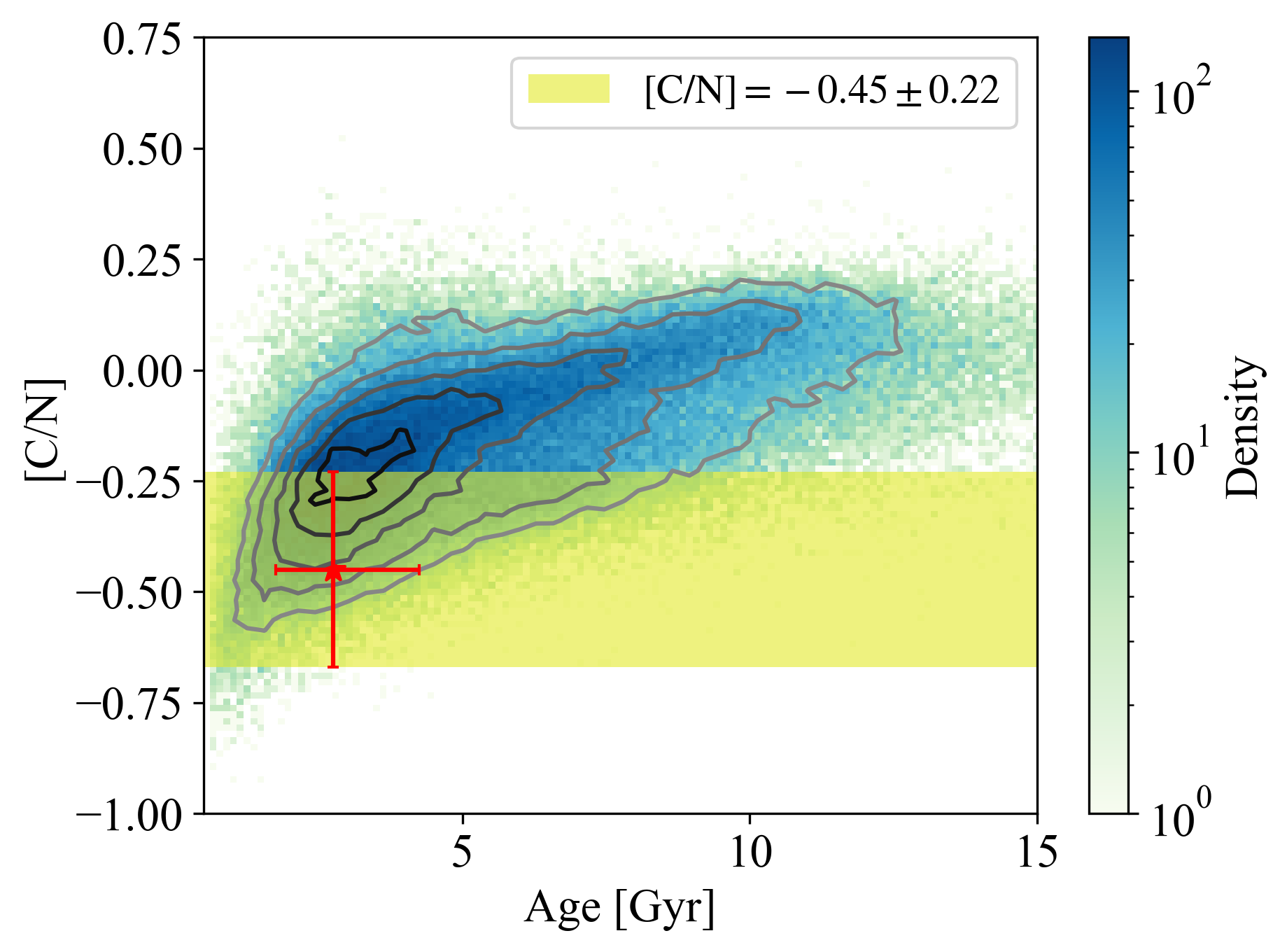}
\includegraphics[scale=0.58]{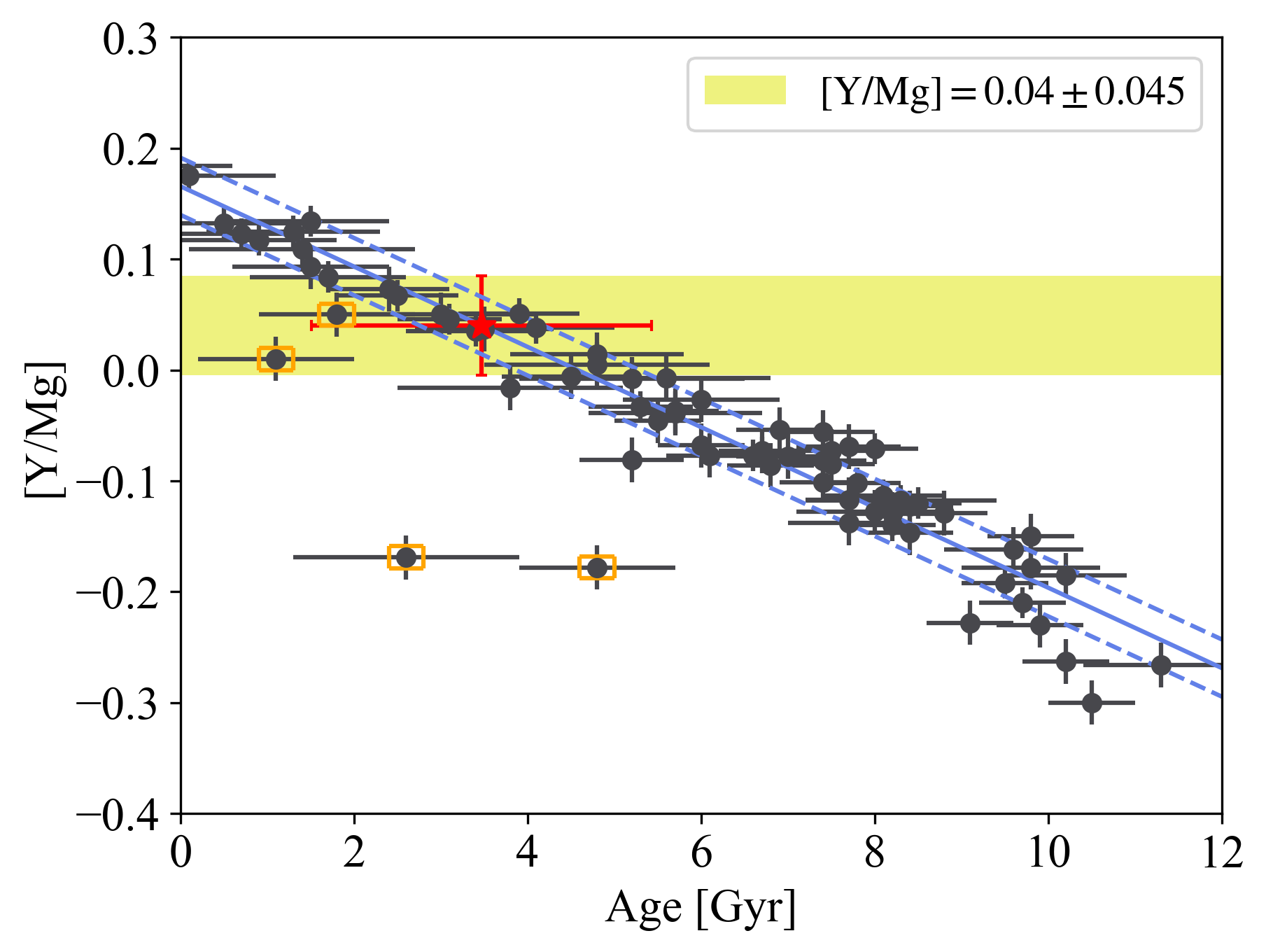}
\caption{Left panel: Density map for age estimates based on [C/N], using red clump stars taken from \citet{2020ApJS..249...29H}.
The yellow horizontal region marks the measurement of [C/N] for 8\,UMi.
The age of 8\,UMi is calculated from the probability distribution of the crossed region;
the red star indicates the final derived age for 8\,UMi.
Right panel: The age-[Y/Mg] relation using nearby 72 nearby stars (black dots) taken from \citet{2020A&A...640A..81N}. 
The blue-solid line and dashed lines are the least-square linear fit and 1-$\sigma$ error for the sample stars. Two binaries and two Na-rich stars are excluded in the fitting (labeled by orange rectangles in the figure; see Sec.\,4.4 of \citealt{2020A&A...640A..81N} for details). 
The yellow horizontal region marks the measurement of [Y/Mg] for 8\,UMi.
The red star represents 8\,UMi, with age given by the linear relation and errors from the uncertainty of the linear relation and measurement errors of [Y/Mg].}
\label{fig:chemical age}
\end{figure}

As mentioned above, the [C/N] ratio at the stellar surface is potentially influenced by the CNO cycle during the first dredge-up, which is correlated with the stellar mass. Thus, [C/N] is a good age indicator for giants because of the mass-age relation in stellar evolution. The left panel of Fig.\,\ref{fig:chemical age} shows the [C/N]-age distribution, using red clump stars again from \citet{2020ApJS..249...29H}. The [Fe/H] range of this sample is from $-$1.0 to +0.5. Their [C/H] and [N/H] are derived from LAMOST low-resolution spectra by adopting a machine learning method using the LAMOST-APOGEE stars in common as a training sample. The typical precision of the derived [C/H] and [N/H] is about 0.1\,dex. We determine $\tau_{\rm [C/N]} = 3.3^{+2.5}_{-1.5}$\,Gyr for 8\,UMi based on the age distribution of red clump stars with [C/N] ranging from $-0.67$ to $-0.23$, the 1-$\sigma$ interval on [C/N] measured for 8\,UMi (see Table\,\ref{Tab:abundance}). 
The systematic error of [C/N] measurement is about 0.15\,dex; its effect on age estimate is around 1.5Gyr.
%The errors in the stellar parameters produce another 0.04\,dex systematic errors on its uncertainty, which is [C/N]$=-0.45 \pm 0.22$ (stat) $\pm \, 0.04$ (sys). This change only increases the upper bound of the age derived from [C/N] by 0.5\,Gyr.

Over the last decade, the ratios between $s$-process and $\alpha$-elements have been found to be chemical clocks for Solar-metallicity stars. Among all the ratios, [Y/Mg]-age relation is tight and applicable for both dwarf and giant stars \citep[e.g.,][]{2020A&A...640A..81N,2021A&A...652A..25C}. 
 By fitting a linear relation\footnote{[Y/Mg] $=$ 0.166 $-$ 0.0362 $\times$ $\tau$.} for age-[Y/Mg] based on a nearby sample\footnote{This sample includes 72 nearby stars with element abundances measured from high resolution and SNR spectra, and age uncertainties between 0.5 and 1.3\,Gyr.} from \citet{2020A&A...640A..81N},  the age inferred from [Y/Mg] for 8\,UMi is $\tau_{\rm [Y/Mg]}=3.5 \pm 2.0$\,Gyr. 
 The uncertainty is given by the 1-$\sigma$ uncertainty of the linear relation and the measurement error of [Y/Mg]. 
 The systematic error, resulting from varying stellar atmospheric parameters and $v_{\rm mic}$ (see Table\,\ref{Tab:abundance}), amounts to 0.09\,dex, corresponding to an uncertainty of 2.5\,Gyr in age.
 Moreover, we investigate the NLTE effects on Mg and Y. In the model atmosphere of 4897/2.60/$-$0.05, the NLTE calculations were conducted for Mg\,I and Y\,II using the model atoms detailed in \citet{2018ApJ...866..153A,2023ApJ...957...10A}. The NLTE abundance corrections, represented as $\mathrm{\Delta_\mathrm{NLTE}}$ = log $\mathrm{\epsilon_{NLTE}}$ $-$ log $\mathrm{\epsilon_{LTE}}$, for the Mg\,I lines at 5528 and 5711\,$\mathrm{\AA}$ are 0.00 and $+$0.03\,dex, respectively. Similarly, for the Y\,II lines at 5289 and 5402\,$\mathrm{\AA}$, the corrections are $+$0.01 and 0.00\,dex, respectively. Thus, the systematic error induced by NLTE effects ($-$0.01\,dex) is negligible comparing to the one caused by errors in the stellar atmospheric parameters and $v_{\rm mic}$.
 %The [Y/Mg]-age relation using high SNR spectra of 72 nearby stars from \citet{2020A&A...640A..81N} is presented in Fig.\,\ref{fig:chemical age}b; the age uncertainty of these 72 stars is typically 0.5-1.3\,Gyr. By fitting a linear relation\footnote{[Y/Mg] $=$ 0.166 $-$ 0.0362 $\times$ $\tau$.} for age-[Y/Mg], 

In summary, the ages of 8\,UMi derived from the two chemical clocks agree very well with each other.
The chemical ages are also consistent with that derived from the kinematic estimate in Sec.\,\ref{3D Space Velocities and Kinematic Age}.

\begin{table*}

\scriptsize
\centering
\caption{Elemental-abundance Ratios of 23 Elements for 8 UMi}\tnote{a}
\begin{threeparttable}
\setlength{\tabcolsep}{2mm}{
\begin{tabular}{crcccccc}
\hline
\hline
\multicolumn{1}{c}{element} & \multicolumn{1}{r}{[X/Fe]} & \multicolumn{1}{c}{$\sigma$\tnote{b}} & \multicolumn{1}{c}{$N$\tnote{c}} & \multicolumn{4}{c}{$\delta$[X/Fe]} \\
& &(dex)& & $\delta$T=$\pm$100K & $\delta$log\,$g$ = $\pm$0.07 & $\delta$[Fe/H]=$\pm$0.07 & $\delta$$v_{\rm mic} = \pm 0.10$\,km\,s$^{-1}$\\
\hline
C & $-$0.50 & 0.10 & -- & $\pm$0.10 & $\pm$0.05 & $\pm$0.10 & $\mp$0.02 \\
N & $-$0.05 & 0.20 & -- & $\pm$0.08 & $\pm$0.02 & $\mp$0.05 & $\mp$0.02 \\
O & $-$0.18 & -- & 1 & $\mp$0.07 & $\pm$0.04 & $\mp$0.05 & 0.00  \\
Na & $+$0.27 & 0.02 & 2 & $\pm$0.08 & 0.00 & $\mp$0.07 & $\mp$0.02 \\
Mg & $+$0.02 & 0.03 & 2 & $\pm$0.07 & $\mp$0.02 & $\mp$0.07 & $\mp$0.03 \\
Al & $+$0.16 & 0.01 & 2 & $\pm$0.07 & 0.00 & $\mp$0.07 & $\mp$0.02 \\
Si & $+$0.02 & 0.01 & 5 & $\mp$0.02 & $\pm$0.01 & $\mp$0.07 & $\mp$0.01 \\
Ca & 0.00 & 0.02 & 6 & $\pm$0.10 & 0.00 & $\mp$0.08 & $\mp$0.04 \\
Sc & $-$0.16 & 0.05 & 5 & $\pm$0.01 & $\pm$0.03 & $\mp$0.07 & $\mp$0.04 \\
Ti & $-$0.01 & 0.05 & 6 & $\pm$0.09 & $\pm$0.01 & $\mp$0.07 & $\mp$0.03 \\
V & $+$0.07 & 0.07 & 5 & $\pm$0.17 & 0.00 & $\mp$0.07 & $\mp$0.03 \\
Cr & $-$0.03 & 0.08 & 4 & $\pm$0.12 & $\pm$0.01 & $\mp$0.08 & $\mp$0.03 \\
Mn & $+$0.07 & 0.04 & 2 & $\pm$0.09 & $\pm$0.01 & $\mp$0.09 & $\mp$0.03 \\
Co & $-$0.03 & 0.05 & 3 & $\pm$0.06 & $\pm$0.01 & $\mp$0.07 & $\mp$0.01 \\
Ni & $-$0.09 & 0.05 & 5 & $\pm$0.05 & $\pm$0.01 & $\mp$0.07 & $\mp$0.03 \\
Zn & $-$0.29 & 0.07 & 1 & $\mp$0.02 & $\pm$0.02 & $\mp$0.07 & $\mp$0.06 \\
Y & $+$0.06 & 0.03 & 2  & 0.00 & $\pm$0.04 & $\mp$0.06 & $\mp$0.02 \\
Zr & 0.00 & 0.07 & 1 & 0.00 & $\pm$0.04 & $\mp$0.07 & $\mp$0.01 \\
Ba & $+$0.25 & 0.06 & 3 & $\pm$0.04 & $\pm$0.01 & $\mp$0.06 & $\mp$0.08 \\
La & $+$0.30 & 0.07 & 1 & $\pm$0.02 & $\pm$0.03 & $\mp$0.05 & $\mp$0.01 \\
Nd & $+$0.31 & 0.04 & 2 & $\pm$0.03 & $\pm$0.03 & $\mp$0.06 & $\mp$0.02 \\
Eu & $+$0.15 & 0.07 & 1 & $\mp$0.01 & $\pm$0.03 & $\mp$0.06 & $\mp$0.01 \\
\hline
A(Li)\tnote{d} & 1.94 & 0.07 & 1 & $\pm$0.15 & 0.00 & $\mp$0.07 & 0.00\\
\hline
\hline
\end{tabular}}
\begin{tablenotes}
\item $^a$: The abundance ratios are derived using the atmospheric parameters: $T_{\rm eff} = 4897 \pm 100 $\,K, log\,$g = 2.60 \pm 0.07$, [Fe/H]\,$= -0.05 \pm 0.07$, $v_{\rm mic} = 1.48 \pm 0.10$\,km\,s$^{-1}$ (see Sec.\,\ref{sec:HRS Observation} for details).
\item $^b$: $\sigma$  represents the statistical uncertainties in measurements. For elements with more than two lines, it is calculated from the scatter of the abundances derived from all lines.
For elements with only one line (Li, Zn, Zr, La, Eu), the statistical error of [Fe/H] is adopted. 
As for C and N, we adopt the 1-$\sigma$ uncertainty as shown in the left panel of Fig.\,\ref{fig:chemical pattern}.
\item $^c$: $N$ represents the number of atomic absorption lines used to derive the chemical abundance.
\item $^d$: We adopted a Solar abundance A (Li)$_{\odot} = 1.01$ taken from \citet{2007SSRv..130..105G}.
\end{tablenotes}
\label{Tab:abundance}
\end{threeparttable}
\end{table*}

\vskip 1cm
\subsection{Is 8\,UMi Formed through a Merger?}   \label{Is 8 UMi formed through merger?}
According to \citet{2023Natur.618..917H}, an old age\footnote{8.6\,Gyr is required in their example binary-evolutionary model.} is required for the binary-merger formation channel of 8\,UMi.
However, the above kinematic and chemical results show that the true age of this system is 3.25--3.50\,Gyr, with a 1-$\sigma$ statistical uncertainty of 1.50 to 3.00\,Gyr. This is consistent with the isochrone-fitting age of $1.9^{+1.2}_{-0.3}$\,Gyr derived under the single-star assumption. 
%Our results thus disfavor the formation of 8\,UMi from the merger between a helium white dwarf (HeWD) star and an red giant branch (RGB) star proposed in \citet{2023Natur.618..917H}.

Another argument that was used in \citet{2023Natur.618..917H} to support the binary-merger scenario is that such a formation channel can explain the Li enhancement of 8\,UMi.
Previous studies have shown that most recognized Li-rich giants are found among the red clump stars \citep[e.g.,][]{2021NatAs...5...86Y}.
To explain this Li enhancement in red clump stars, one possible scenario is the merger of a HeWD and an RGB that triggers a convection shell to synthesize Li, leading to Li-rich red clump stars \citep{2020ApJ...889...33Z}.
%This is also the model adopted by \citet{2023Natur.618..917H} to form 8\,UMi (classified as a red clump from asteroseismology), and thus avoid the engulfment of the close-in planet Halla.
%In \citet{2023Natur.618..917H}, the detection of Li enhancement in  8\,UMi is thus thought to provide indirect evidence for the binary-merger channel. 
However, other competing models that do not invoke binary mergers can also produce Li-rich red clump stars, such as planet engulfment \citep{2016ApJ...833L..24A} and the lithium enhancement in the \text{tip of the RGB} process or in the helium flash \citep{2014ApJ...784L..16S}. Given that there is another giant planet with a relatively wide orbit in 8 UMi, the chance is high that one or several small planets may have existed on inner orbits \citep{2021ARA&A..59..291Z}, and the ingestion of such close-in planets would be sufficient to explain the observed Li abundance \citep[e.g.,][]{2016ApJ...833L..24A}.

In summary, the binary-merger formation channel of 8\,UMi proposed by \citet{2023Natur.618..917H} is not supported by our kinematic and chemical analysis of 8\,UMi. Alternative models are required to explain the existence of the close-in planet Halla orbiting 8\,UMi.

\vskip 1.5cm
\section{Conclusions} \label{sec:conclusion}

Using astrometric data from {\it Gaia} DR3 and a high-resolution spectrum from the SOPHIE archive, we determine the mass, age, 3D position, space velocity, and the chemical-abundance pattern for 8\,UMi. Our results reveal that the planet's host star 8\,UMi is a single middle-aged thin-disk star with a normal orbital rotation speed of $V_{\phi} = 244.96$\,km\,s$^{-1}$. 
It also exhibits a Solar-like metal abundance and [$\alpha$/Fe] ratio. 
To test the binary-merger formation scenario for 8\,UMi, we have determined its age by different methods, including isochrone-fitting assuming a single-star evolution, and kinematic and chemical age estimates from well-established age--space velocities/chemical element-abundances relations. 
The four derived ages are:

1) Isochrone age: $\mathrm{\tau_{isochrone}=1.9^{+1.2}_{-0.3} \thinspace Gyr}$;

2) Kinematic age: $\mathrm{\tau_{kinematic}=3.5^{+3.0}_{-2.0} \thinspace Gyr}$;

3) [C/N] age: $\mathrm{\tau_{[C/N]}=3.3^{+2.5}_{-1.5} \thinspace Gyr}$;

4) [Y/Mg] age: $\mathrm{\tau_{[Y/Mg]}=3.5 \pm 2.0 \thinspace Gyr}$.

These age measurements consistently suggest that 8 UMi is a middle-aged star formed approximately 2--4 Gyr ago. This is much younger than the timescale that is needed to go though the binary-merger process of \citet{2023Natur.618..917H}. Further investigations are needed to fully resolve the puzzle of the 8 UMi planetary system.

\begin{acknowledgements}

$\it{Acknowledgements} \,$ This work is funded by the National Key R\&D Program of China (No. 2019YFA0405500)
and the National Natural Science Foundation of China (NSFC grant Nos. 12090040, 12090044, 11833006, 12133005, 12173021 and 11833002). T.C.B. 
acknowledges partial support for this work from grant PHY 14-30152; Physics Frontier Center/JINA Center for the Evolution of the Elements (JINA-CEE), and OISE-1927130: The International Research Network for Nuclear Astrophysics (IReNA), awarded by the US National Science Foundation. 
W.Z. is also supported by the CASSACA grant CCJRF2105.

Thanks to the maintenance of the SOPHIE data archive by Haute Provence Observatory (\url{http://atlas/obs-hp.fr/sophie/}), we 
were able to obtain access to the high-resolution spectrum used in this study. We also used data from the European Space Agency mission Gaia (\url{https://www.cosmos.esa.int/Gaia}), processed by the Gaia Data Processing and Analysis Consortium (DPAC; see \url{http://www.cosmos.esa.int/web/Gaia/dpac/consortium}). 
Besides, this research has made use of the NASA Exoplanet Archive (\url{https://exoplanetarchive.ipac.caltech.edu/}), which is operated by the California Institute of Technology, under contract with the National Aeronautics and Space Administration under the Exoplanet Exploration Program.

\end{acknowledgements}

\clearpage
\bibliography{sophie_2930170}{}
\bibliographystyle{aasjournal}

\appendix

\section{Line data and synthetic fitting examples in the solar spectrum} \label{Line data}
\setcounter{table}{0}
\renewcommand{\thetable}{A.\arabic{table}}

\begin{longtable}{ccccccc}
\caption{Line Data Used in the Abundance Derivation} \\
\hline
\hline
Atom & $\lambda$ & $E_{\mathrm{exc}}$ & $\log gf$ & EW$_{\odot}$ & EW$_{\mathrm{8\,UMi}}$ & [X/Fe]$\mathrm{_{8\,UMi}}$ \\
     & ($\mathrm{\AA}$) & (eV) & & ($\mathrm{m\AA}$) & ($\mathrm{m\AA}$) &  \\
\hline
\endfirsthead

\multicolumn{7}{c}%
{{\bfseries \tablename \, \thetable{.}} -- continued} \\
\hline
\hline
Atom & $\lambda$ & $E_{\mathrm{exc}}$ & $\log gf$ & EW$_{\odot}$ & EW$_{\mathrm{8\,UMi}}$ & [X/Fe]$\mathrm{_{8\,UMi}}$ \\
     & ($\mathrm{\AA}$) & (eV) & & ($\mathrm{m\AA}$) & ($\mathrm{m\AA}$) & (dex) \\
\hline
\endhead

\hline 
\endfoot

\hline
\endlastfoot

Li	&	6707.80 	&	0.17 	&	0.00 	&	--	&	138.11 	&	+0.98 	\\
O	&	6300.30 	&	0.00 	&	$-$9.58 	&	7.65 	&	32.16 	&	$-$0.18 	\\
Na	&	6154.23 	&	2.10 	&	$-$1.58 	&	37.92 	&	72.79 	&	+0.28 	\\
Na	&	6160.75 	&	2.10 	&	$-$1.26 	&	59.40 	&	95.42 	&	+0.25 	\\
Mg	&	5528.00 	&	4.34 	&	$-$0.47 	&	237.72 	&	227.83 	&	+0.04 	\\
Mg	&	5711.09 	&	4.34 	&	$-$1.70 	&	107.42 	&	130.84 	&	0.00 	\\
Al	&	6696.02 	&	3.14 	&	$-$1.55 	&	38.00 	&	66.36 	&	+0.16 	\\
Al	&	6698.67 	&	3.14 	&	$-$1.90 	&	21.84 	&	45.59 	&	+0.18 	\\
Si	&	6142.48 	&	5.62 	&	$-$1.50 	&	34.14 	&	42.52 	&	+0.01 	\\
Si	&	6145.02 	&	5.62 	&	$-$1.42 	&	38.89 	&	48.10 	&	+0.02 	\\
Si	&	6237.32 	&	5.61 	&	$-$1.10 	&	60.06 	&	70.13 	&	+0.02 	\\
Si	&	6243.81 	&	5.62 	&	$-$1.27 	&	46.56 	&	56.90 	&	+0.02 	\\
Si	&	6244.47 	&	5.62 	&	$-$1.29 	&	46.69 	&	58.14 	&	+0.02 	\\
Ca	&	6156.02 	&	2.52 	&	$-$2.46 	&	10.96 	&	28.72 	&	+0.01 	\\
Ca	&	6161.30 	&	2.51 	&	$-$1.29 	&	65.28 	&	103.61 	&	+0.03 	\\
Ca	&	6166.44 	&	2.51 	&	$-$1.14 	&	71.15 	&	103.04 	&	$-$0.02 	\\
Ca	&	6169.04 	&	2.51 	&	$-$0.80 	&	91.31 	&	125.18 	&	$-$0.02 	\\
Ca	&	6455.60 	&	2.51 	&	$-$1.35 	&	57.00 	&	94.71 	&	+0.02 	\\
Ca	&	6499.65 	&	2.51 	&	$-$0.82 	&	83.98 	&	120.18 	&	$-$0.03 	\\
ScII	&	5526.81 	&	1.76 	&	0.02 	&	71.97 	&	115.54 	&	$-$0.18 	\\
ScII	&	5640.97 	&	1.49 	&	$-$0.96 	&	40.38 	&	85.80 	&	$-$0.08 	\\
ScII	&	5657.87 	&	1.50 	&	$-$0.40 	&	67.21 	&	108.67 	&	$-$0.13 	\\
ScII	&	5669.03 	&	1.49 	&	$-$1.05 	&	35.00 	&	71.40 	&	$-$0.22 	\\
ScII	&	6245.63 	&	1.51 	&	$-$1.06 	&	35.52 	&	72.53 	&	$-$0.17 	\\
TiII	&	5211.53 	&	2.59 	&	$-$1.45 	&	33.57 	&	59.91 	&	$-$0.03 	\\
TiI	&	5219.70 	&	0.02 	&	$-$2.24 	&	28.17 	&	98.08 	&	+0.02 	\\
TiII	&	5418.77 	&	1.58 	&	$-$2.07 	&	49.77 	&	83.76 	&	$-$0.10 	\\
TiI	&	5648.57 	&	2.50 	&	$-$0.32 	&	10.89 	&	38.58 	&	+0.03 	\\
TiI	&	5662.16 	&	2.32 	&	$-$0.10 	&	23.74 	&	63.38 	&	 0.00 	\\
TiI	&	6126.22 	&	1.07 	&	$-$1.38 	&	22.67 	&	83.01 	&	 0.00 	\\
V	&	5670.85 	&	1.08 	&	$-$0.45 	&	21.11 	&	84.83 	&	+0.18 	\\
V	&	5737.07 	&	1.06 	&	$-$0.70 	&	14.09 	&	64.21 	&	+0.03 	\\
V	&	6081.45 	&	1.05 	&   $-$0.70 	&	14.88 	&	69.13 	&	+0.03   \\
V	&	6274.66 	&	0.27 	&	$-$1.75 	&	9.35 	&	64.19 	&	+0.03 	\\
V	&	6285.17 	&	0.28 	&	$-$1.65 	&	10.83 	&	70.82 	&	+0.06 	\\
Cr	&	5287.18 	&	3.44 	&	$-$0.95 	&	11.36 	&	34.10 	&	+0.10 	\\
Cr	&	5300.75 	&	0.98 	&	$-$2.09 	&	59.85 	&	113.35 	&	$-$0.07 	\\
Cr	&	5783.07 	&	3.32 	&	$-$0.42 	&	34.46 	&	60.41 	&	$-$0.06 	\\
Cr	&	5787.92 	&	3.32 	&	$-$0.15 	&	48.21 	&	78.27 	&	$-$0.07 	\\
Mn	&	5399.47 	&	3.85 	&	$-$0.18 	&	40.98 	&	76.07 	&	+0.04 	\\
Mn	&	5413.67 	&	3.86 	&	$-$0.65 	&	25.25 	&	52.05 	&	+0.09 	\\
Co	&	5352.05 	&	3.58 	&	0.01 	&	25.00 	&	58.31 	&	0.00 	\\
Co	&	5359.20 	&	4.15 	&	0.04 	&	9.67 	&	22.59 	&	$-$0.08 	\\
Co	&	5647.24 	&	2.28 	&	$-$1.55 	&	15.11 	&	56.60 	&	0.00 	\\
Ni	&	5625.32 	&	4.09 	&	$-$0.70 	&	39.76 	&	62.57 	&	$-$0.10 	\\
Ni	&	5643.08 	&	4.17 	&	$-$1.24 	&	16.63 	&	31.66 	&	$-$0.10 	\\
Ni	&	6086.29 	&	4.27 	&	$-$0.47 	&	43.90 	&	62.28 	&	$-$0.16 	\\
Ni	&	6108.12 	&	1.68 	&	$-$2.51 	&	67.40 	&	121.27 	&	$-$0.09 	\\
Ni	&	6111.08 	&	4.09 	&	$-$0.82 	&	34.91 	&	59.74 	&	$-$0.02 	\\
Zn	&	4810.53 	&	4.08 	&	$-$0.22 	&	76.37 	&	81.34 	&	$-$0.29 	\\
Y	&	5289.82 	&	1.03 	&	$-$1.88 	&	4.25 	&	23.86 	&	+0.08 	\\
Y	&	5402.77 	&	1.84 	&	$-$0.58 	&	12.65 	&	40.71 	&	+0.04 	\\
Zr	&	5112.27 	&	1.67 	&	$-$0.85 	&	9.30 	&	36.58 	&	0.00 	\\
Ba	&	5853.67 	&	0.60 	&	$-$1.00 	&	63.13 	&	119.69 	&	+0.18 	\\
Ba	&	6141.71 	&	0.70 	&	$-$0.08 	&	114.63 	&	117.42 	&	+0.28 	\\
Ba	&	6496.90 	&	0.60 	&	$-$0.38 	&	97.75 	&	167.73 	&	+0.28 	\\
La	&	6390.48 	&	0.32 	&	$-$1.46 	&	3.31 	&	30.67 	&	+0.30 	\\
Nd	&	5311.45 	&	0.99 	&	$-$0.53 	&	3.15 	&	25.79 	&	+0.28 	\\
Nd	&	5319.82 	&	0.55 	&	$-$0.34 	&	12.00 	&	60.83 	&	+0.33 	\\
Eu	&	6645.10 	&	1.38 	&	0.22 	&	6.14 	&	28.67 	&	+0.15 	

\label{Tab:line data}
\end{longtable}

\newpage

\setcounter{figure}{0}
\renewcommand{\thefigure}{A.\arabic{figure}}

\begin{figure}[ht!]
\centering
\includegraphics[scale=0.18]{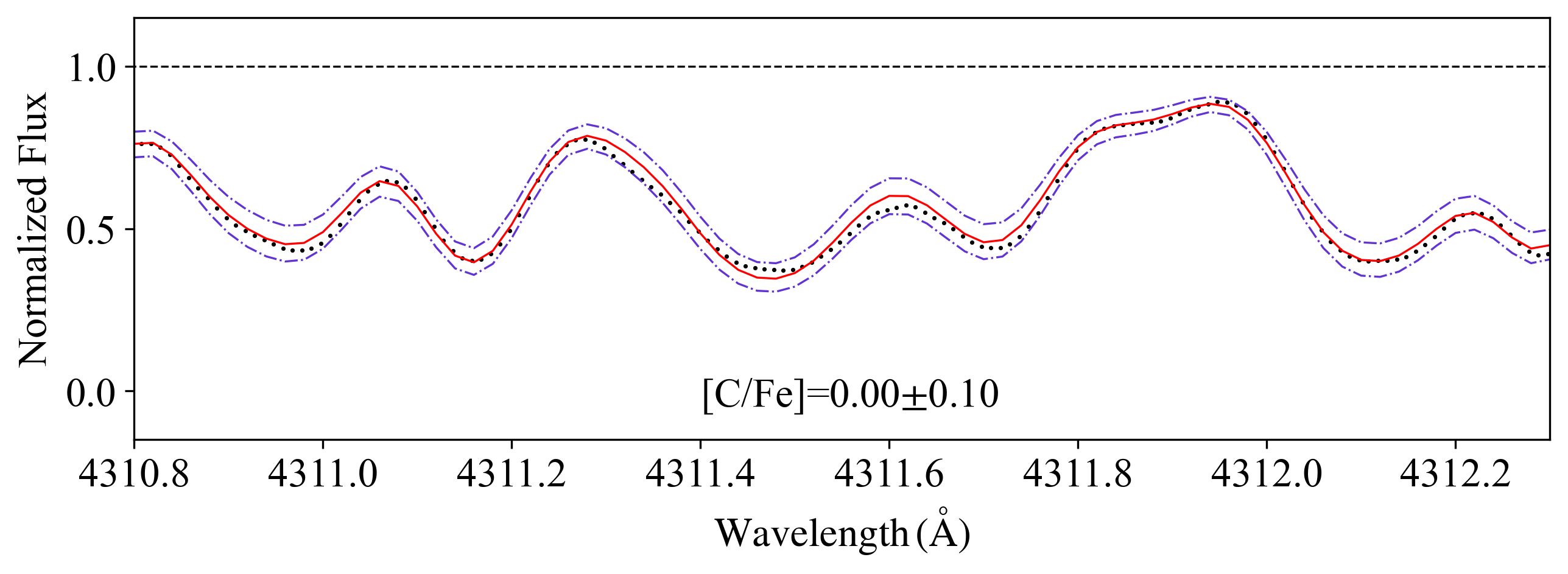}
\includegraphics[scale=0.18]{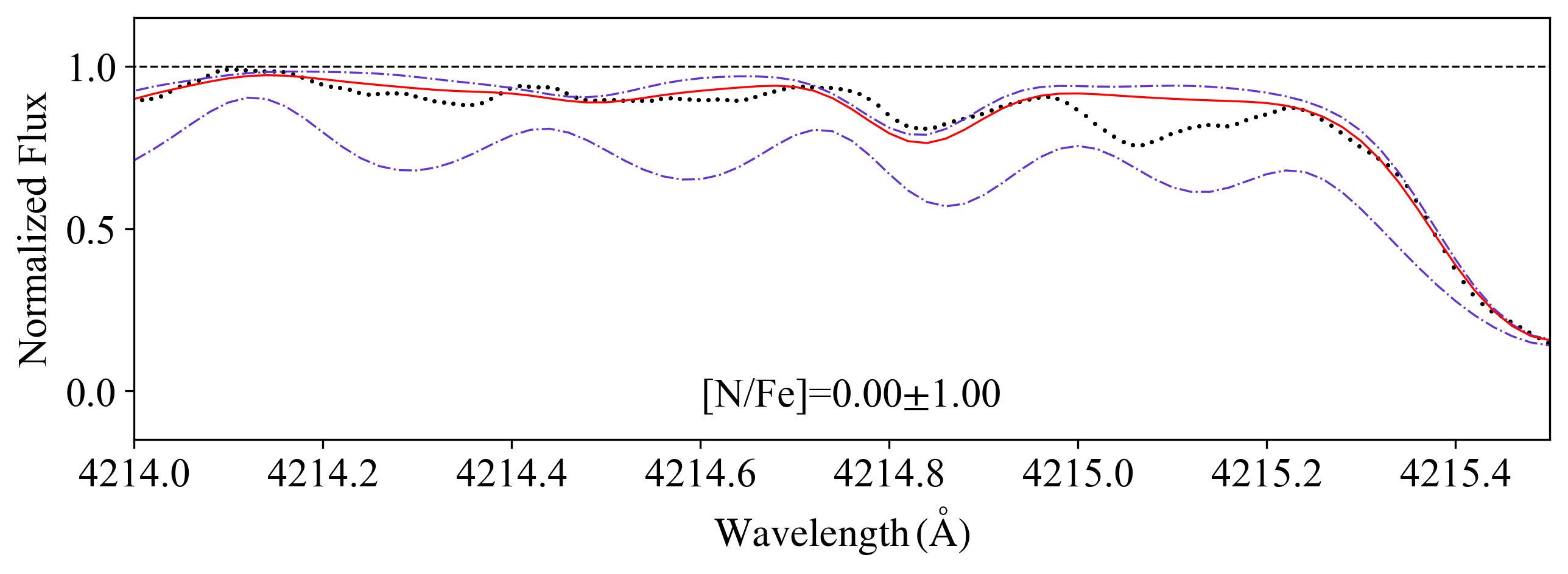}
\includegraphics[scale=0.12]{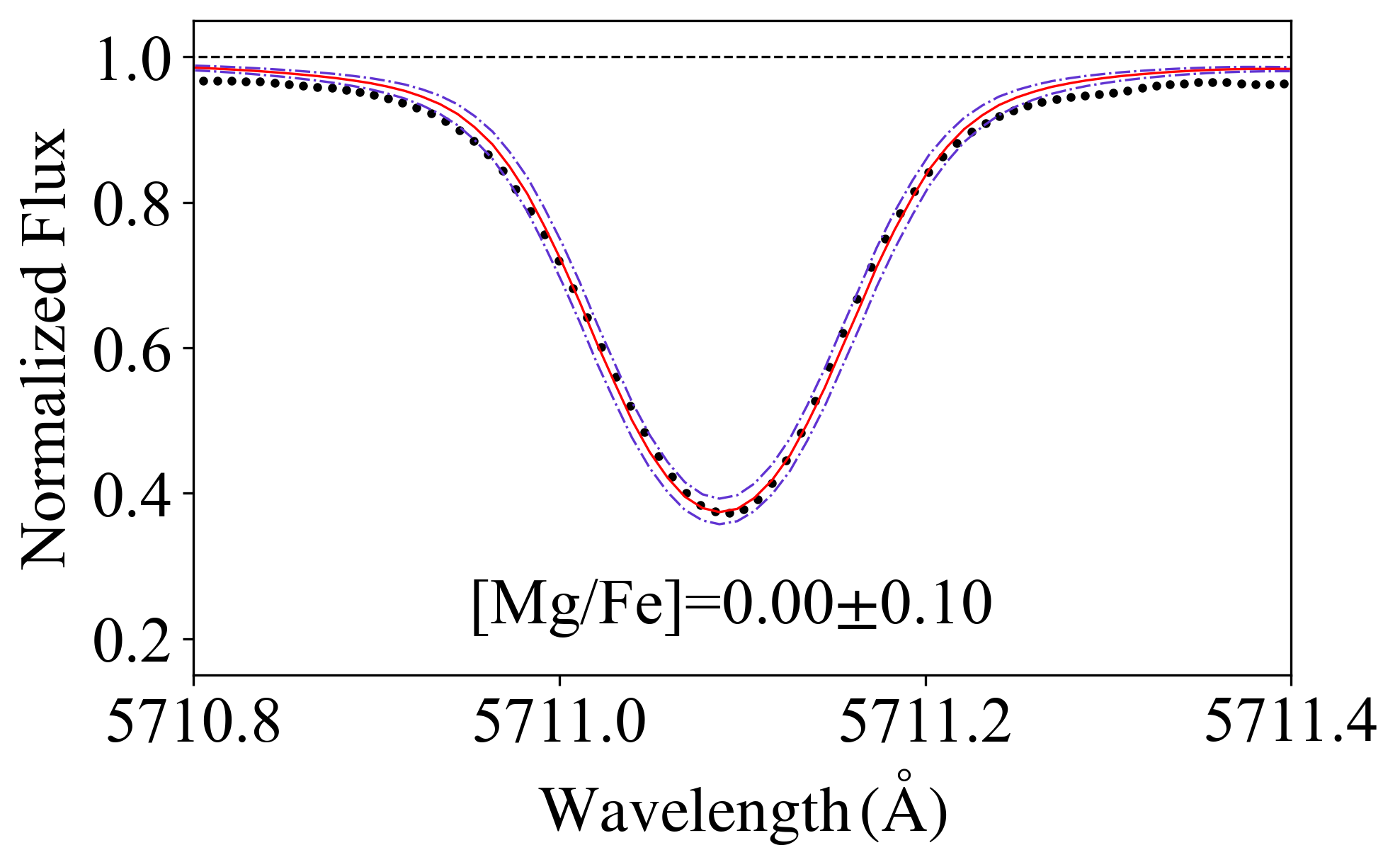}
\includegraphics[scale=0.12]{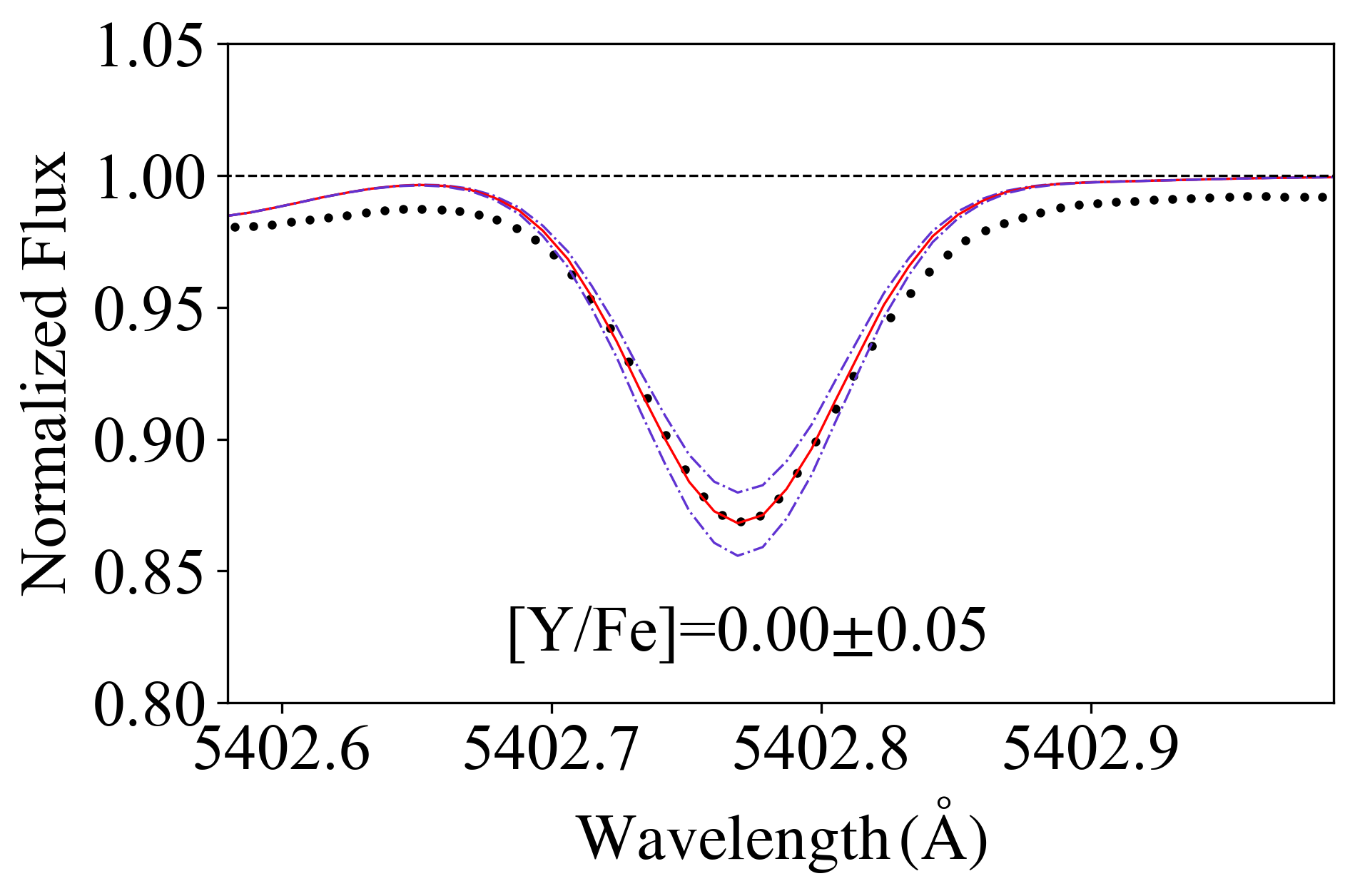}
\caption{The synthetic line-fitting results (red solid line) of CH, CN, Mg, and Y in the observed Solar spectrum (black dots). The Solar atmospheric parameters are set to $T_{\rm eff} = 5777 $\,K, log\,$g = 4.44 $, [Fe/H]\,$= 0.00 $, and $v_{\rm mic} = 1.00 $\,km\,s$^{-1}$. The upper two panels show the CH and CN band with the observed Solar spectrum taken by SOPHIE under HR mode, with abundance variations of 0.10 and 1.00\,dex, respectively. Since molecular lines differ between dwarf and giant stars, the CN band is not well-reproduced in the Solar spectrum. The lower two panels show the fitting of Mg 5711\,$\mathrm{\AA}$ and Y 5402\,$\mathrm{\AA}$ lines with abundance variations of 0.10 and 0.05\,dex, respectively.}
\label{fig:solar_fitting}
\end{figure}

\end{document}